# Reversible Magneto-ionic Modification of Metallic Magnetic Thin Films

Md Golam Hafiz,* Hari Babu Vasili, Philippa Shepley, Mannan Ali, Andrew J. Britton, Oscar Cespedes, Weibin Li, Manuel Valvidares, Rohit Pachat, Wilfried Fotso, Mourad Cherif, Shimpei Ono, Yves Roussigne, Mohamed Belmeguenai, and Gavin Burnell*

**ABSTRACT:** Magnetic multilayers with structural inversion asymmetry, perpendicular magnetic anisotropy (PMA), and large Dzyaloshinskii-Moriya interaction (DMI) are widely studied for magnetic domain wall- and skyrmion-based data processing applications. In this work, we report a controlled, nonvolatile, but reversible local modification of such a system by electrically driven oxygen migration through the metallic film structure by using ionic liquid gating. Our findings provide direct evidence for the modification of the oxidation state at the heavy metal (HM)/ferromagnet (FM) interface. As a result, we observed changes in fundamental magnetic properties, such as an increase in the effective anisotropy constant ($K_{eff}$) and coercive field ($H_c$) with oxidation, and a decrease with the reduction of oxygen ions. Positive gate voltages relative to the thin film extract oxygen ions from the magnetic layer, significantly reducing the domain nucleation field. In contrast, negative voltages drive oxygen ions into the magnetic layer, leading to a reversible increase in the extent of domain wall pinning near saturation. We observe a corresponding decrease in the magnetic moment and DMI, with negative voltage reducing both. This magneto-ionic modulation in fully metallic structures is beneficial for spintronic device applications, particularly in field-programmable domain wall and skyrmion devices.

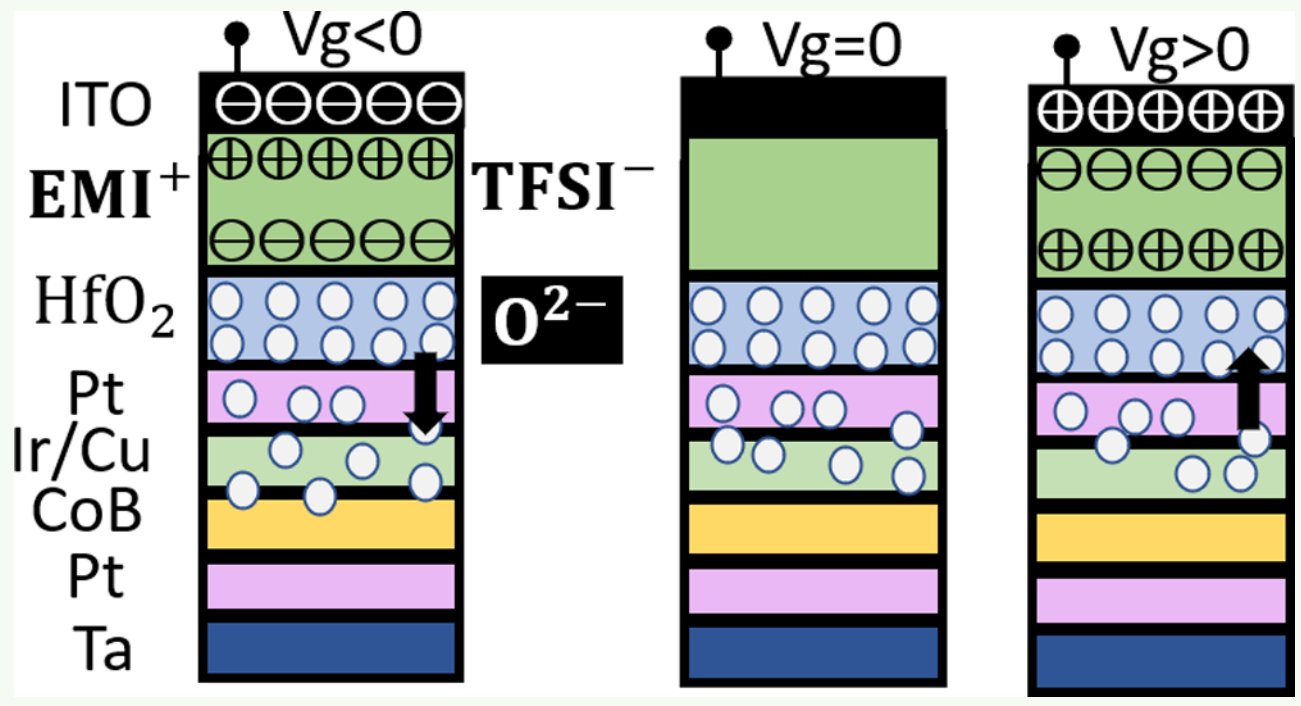

## ■ INTRODUCTION

Magnetic properties of a thin film multilayer strongly depend on the material structure and the quality of the interface. The possibility of controlling the structure and interface is of great importance in the development of compact, fast, and low-power electronic devices. Electric (E)-field-induced migration of ions such as oxygen or hydrogen is already used as a technique to manipulate magnetic properties.[1−3] This so-called magneto-ionic modulation offers nonvolatile effects[2,4] at room temperature and is of great interest due to its capability to modify PMA by up to approximately 5 pJ/Vm.[1,3] Although recent studies have already demonstrated E-field-induced magnetoionic modification of magnetic moment,[5,6] anisotropy,[7] exchange bias,[8−10] DMI,[11] domain wall (DW) properties,[4,12,13] and skyrmion properties,[13] most of the research has focused on investigating the magnetic layer adjacent to an oxide layer. However, PMA structures often include metallic layers on both sides of the ferromagnetic layer.[14−16] For example, HM1/FM/HM2 structures demonstrate high PMA and DMI due to spin orbit coupling (SOC) and intricate hybridization of orbitals at the interfaces.[17,18] These features are particularly valuable in domain wall and skyrmion-based memory devices,[14,19] where the ability to control magnetic properties by modulating the interfacial structure and chemistry enables significant advancements in reconfigurable spintronic memory devices. The literature has also reported the controlled oxidation of heavy metal (HM) layers and its effects on magnetic properties. For example, Nath et al. (2022) investigated the gradual oxidation of Pt and its impact on the spin−orbit torque in Co/Pt systems.[20] Similarly, Xie et al. (2019) studied the effect of Pt oxidation on spin−orbit torque in the Pt/Co system.[21]

In this work, we have developed ultrathin layers consisting of Ta (10 Å)/Pt (23 Å)/ $Co_{68\%}B_{32\%}$ (x Å)/Ir or Cu (2 Å)/Pt (y

**Received:** February 3, 2026
**Revised:** March 12, 2026
**Accepted:** March 14, 2026
**Published:** March 27, 2026

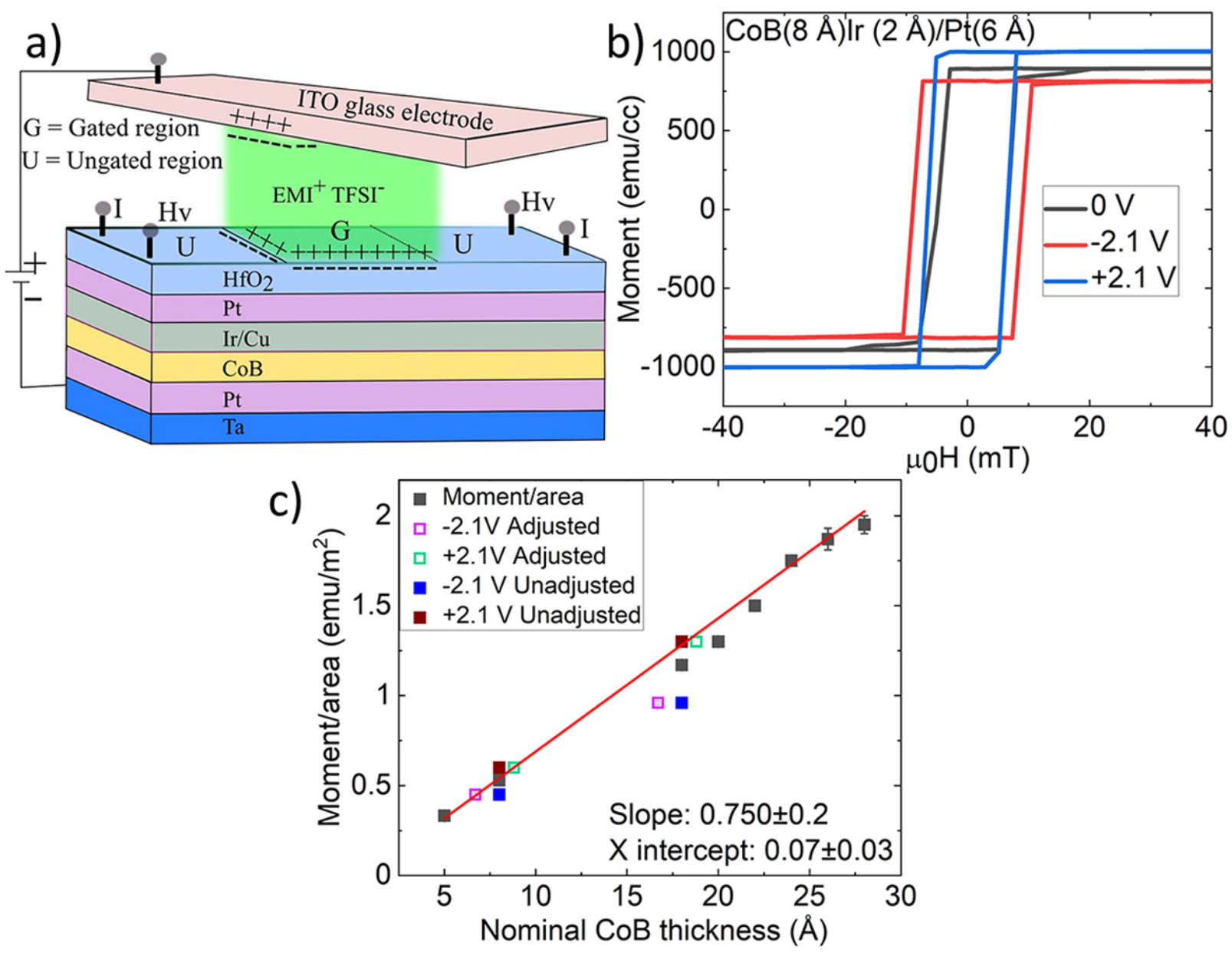

**Figure 1.** (a) Ionic liquid $EMI^+$ $TFSI^-$ is applied on the samples (Ta (10 Å)/Pt (23 Å)/$Co_{68\%}B_{32\%}$ (8 Å)/Ir (2 Å)/Pt (6 Å)/$HfO_2$ (25 Å) and Ta (10 Å)/Pt (23 Å)/$Co_{68\%}B_{32\%}$ (8 Å)/Cu (2 Å)/Pt (6 Å)/$HfO_2$ (25 Å)), and an ITO glass is used as a top electrode. (b) Out of plane hysteresis loop shows that the coercive field is changing with applied voltages. (c) The $M_s$ is calculated from magnetic moment per unit area versus the film thickness of as grown samples. The blue and wine points represent gated $M_s$ values. The gated values are shifted (hollow points) to align with the straight line fit, with the shift from the original positions calculated using the toy model presented in Supporting Information, Section 2.

Å)/$HfO_2$ (25 Å), where x = 5 to 30 and y = 2 to 8. Most of the data presented here are from CoB (8 Å) to Pt (6 Å), where exceptions were specified. In ultrathin multilayer structures, CoB is amorphous and exhibits strong thermal stability as well as strong interfacial interactions. It also contributes to the stabilization of small, circular, and uniform magnetic domains, such as bubble domains. Therefore, it is suited for domain wall or skyrmion-based memory devices.[22−26] The metallic layers were deposited by sputtering, while the $HfO_2$ was grown using ex-situ atomic layer deposition. Due to the ultrathin top Pt layer, oxygen may penetrate the system. Although our growth system is capable of forming continuous Pt films at thicknesses of approximately 4−5 Å, minor discontinuities at this scale cannot be excluded. We have demonstrated that E-field-induced oxygen migration can be used to modify the interfacial chemistry of the system. This, in turn, alters the magnetic properties of the system, including saturation magnetization ($M_s$) and effective DMI ($DMI_{eff}$). We observed the chemical state of the interfaces using X-ray magnetic circular dichroism (XMCD) and X-ray absorption spectroscopy (XAS).

## ■ RESULTS AND DISCUSSION

### Sample Structure and Magnetic Behavior

Figure 1(a) shows the sample structure and electrical connections.

The gate voltage is applied between the film, which acts as a bottom electrode, and an indium tin oxide (ITO) coated glass substrate, serving as a top electrode. Ionic liquid [$EMI^+$]-[$TFSI^-$], or 1-ethyl-3-methylimidazolium bis(trifluoromethanesulfonyl) imide, was used to apply voltage to all samples presented in this work. For gating, a sample with dimensions of approximately 10 mm × 5 mm was used. A small drop of the ionic liquid was placed in the center of the sample, and an ITO glass electrode (approximately 5 mm in width) was positioned on top of the liquid. Due to capillary action, the ionic liquid spreads uniformly beneath the ITO electrode, resulting in a gated area of around 25 $mm^2$. All measurements were performed after the gate voltage was turned off. When a gate voltage is applied, ions within the liquid gate move toward the opposite polarity, forming electric double layers. The distance between two positive and negative ions in a double layer is ≈1 nm,[2,27] creating a nanocapacitor with a large capacitance of up to 170 $\mu F/cm^2$. Therefore, a small applied voltage produces a large E-field effect, altering the carrier density by an order of magnitude between $10^{14}/cm^2$ to $10^{15}/cm^2$.[28]

Recent studies show that the E-field moves mobile oxygen from the oxide layer in a magnetic multilayer system.[2,29−32] Therefore, we interpret our results as an effect of oxygen migration. The oxygen ions move toward the CoB layer with an applied negative voltage, which changes the oxidation state of the sample. It is worth mentioning that the anisotropy changes with voltage (discussed in more detail below); however, it consistently remains in the PMA regime (Figure 1(b, c)). The $H_c$ changes with the applied voltage in the samples, as also observed in the Kerr microscope and XMCD hysteresis loops (Supporting Information, Figure S1 and Figure S2). Li et al. (2017)[33] studied the Co/SrCoO interface and observed a change in coercive field similar to our findings. They attributed this change to Co-3d and the O-2p hybridization.

With an applied negative voltage, the moment per unit area decreases, while with a positive voltage, it increases. The $M_s$ is estimated using a linear fit to a data set of moment per unit area versus film thickness prior to gating. This yielded an $M_s$ of 750 ± 20 emu/cc, with an effective magnetic dead layer of 0.7 Å. Satchell et al. (2021)[34] found an $M_s$ of 760 ± 90 emu/cc, where the Boron concentration was also 32%. They also observe the polarization of the Pt layer, where in our system, with a significantly thinner Pt capping layer, we observed a dead layer. Tan et al. (2021)[24] reported a dead layer of 4 Å to

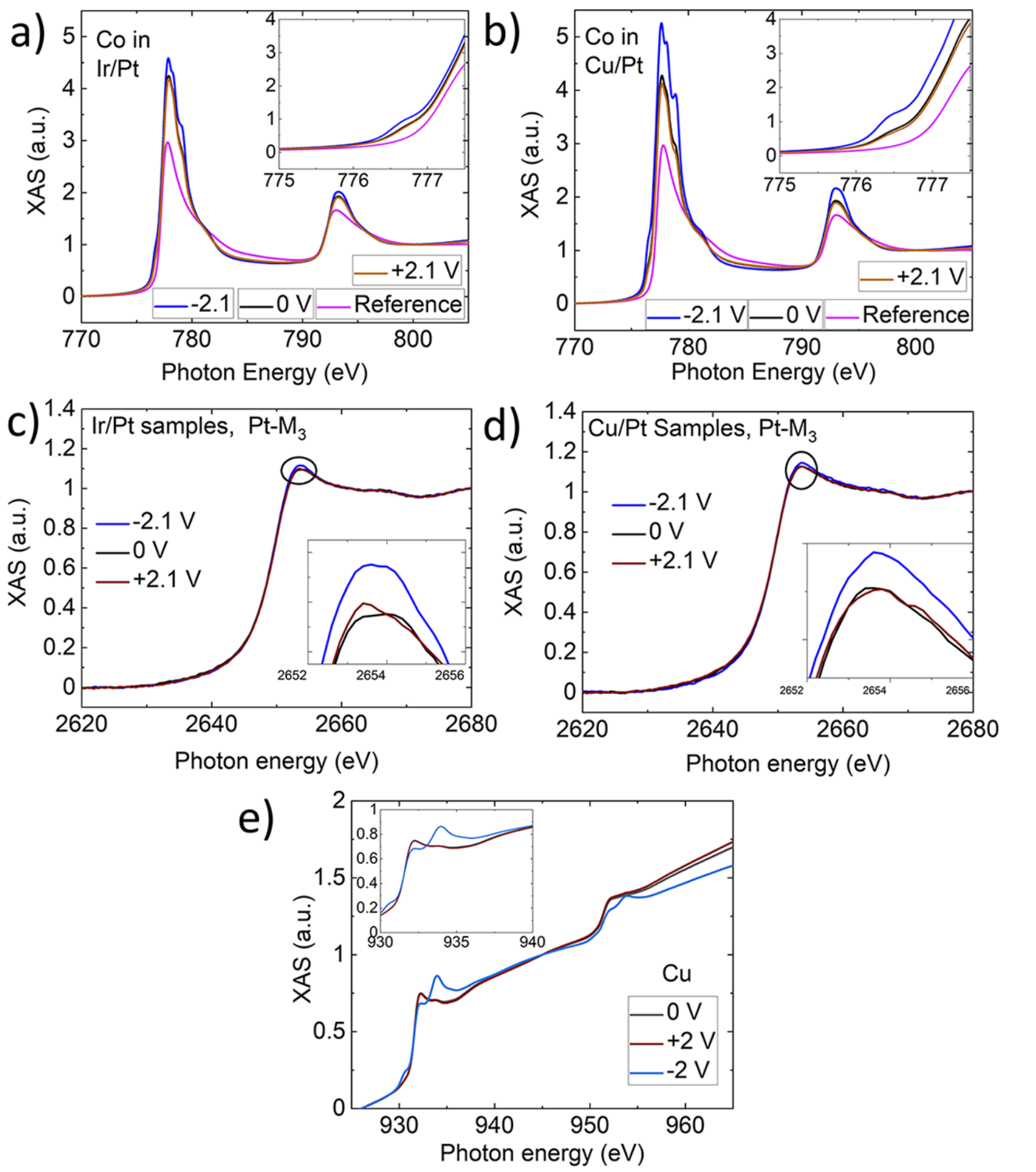

**Figure 2.** XAS spectra of (a) Ir/Pt and (b) Cu/Pt samples, showing a negative voltage increasing the Co−$L_{2,3}$ prepeak, indicating CoO formation. A reference sample Ta (14 Å)/Pt (23 Å)/CoB (8 Å)/Ir (2 Å)/Pt (6 Å)/Ta (10 Å) is used to compare the Co−$L_{2,3}$ prepeak of samples studied in this work. The Ta cap protects the sample from oxidation, indicating that the as-grown samples underwent oxidation between dc magnetron sputtering and ex-situ ALD. XAS spectra at the Pt−$M_3$ absorption edge of (c) Ir/Pt samples and (d) Cu/Pt samples, indicating that oxidation is higher for negatively gated samples compared to as-grown and positively gated samples. (e) XAS spectra of Cu (in Cu/Pt sample), also showing the high oxygen concentration with a negative voltage.

6 Å in CoB/HM systems. We employ a simple model (see Supporting Information, Figure S3) that treats the effect of gating in a sample as increasing (negative gate voltage) or reducing (positive gate voltage) a magnetically dead layer from the 0.7 Å in the ungated samples. From this we infer a total change in thickness of the dead layer by 1.5 ± 0.1 Å from positive gating to negative gating. Although this simple model implies the complete reduction of oxidized Co by a positive gate, it does not take into account the effect of any magnetic proximity effect with Pt.

## X-ray Absorption Spectroscopy (XAS) Based Reconstruction of Oxidation States and Local Structural Environment

In order to understand the electronic structure and chemical reconstruction of the materials, we studied the samples by using XAS and XMCD measurements. Figure 2 presents the Co−$L_{2,3}$ XAS spectra of the Ir/Pt sample (a) and the Cu/Pt sample (b). A controlled multilayer sample with a Ta cap was used as a reference, where the Ta layer prevents oxidation. The XAS of the ungated (0 V) samples compared to the reference Co−$L_{2,3}$ spectra provides evidence that Co is partially oxidized between the two growth steps (sputtering and ex situ ALD). This is further confirmed by observing a larger whiteline intensity and a prepeak shoulder at approximately 776.5 eV (insets), before the Co metallic peaks, associated with $Co^{2+}$.[35] When a negative voltage (−2.1 V) is applied, the spectral features change significantly, indicating an increased level of oxidation in the CoB layer. In contrast, applying a positive voltage (+2.1 V) causes only a slight change in the oxidation level, which remains nearly the same as that in the ungated state (0 V). X-ray photoelectron spectroscopy (XPS) data (Supporting Information, Figure S4) on the Co-L absorption edges is in good agreement with the oxidation observed in the as-grown samples. Under a negative voltage, the oxide peak increases, suppressing the Co metal peak. In contrast, under positive voltage, the metal peak increases, suppressing the oxidation of CoB. Both the Co-L XAS and XPS data clearly demonstrate a migration of oxygen ions to (or from) the CoB layer under applied voltages. A similar reversible effect is observed on the absorption edges of Pt-$M_3$ and Cu-$L_{2,3}$, as discussed in the following.

Figures 2(c, d) show the Pt-$M_3$ XAS of the Ir/Pt and Cu/Pt samples, respectively. The Pt $M_3$ whiteline intensity peak indicates oxidation behavior in the as-grown samples.[36] A negative voltage significantly increases oxidation, resulting in a higher intensity of the Pt white line, suggesting that oxygen from $HfO_2$ migrates toward the magnetic layer.[2,11,29] In

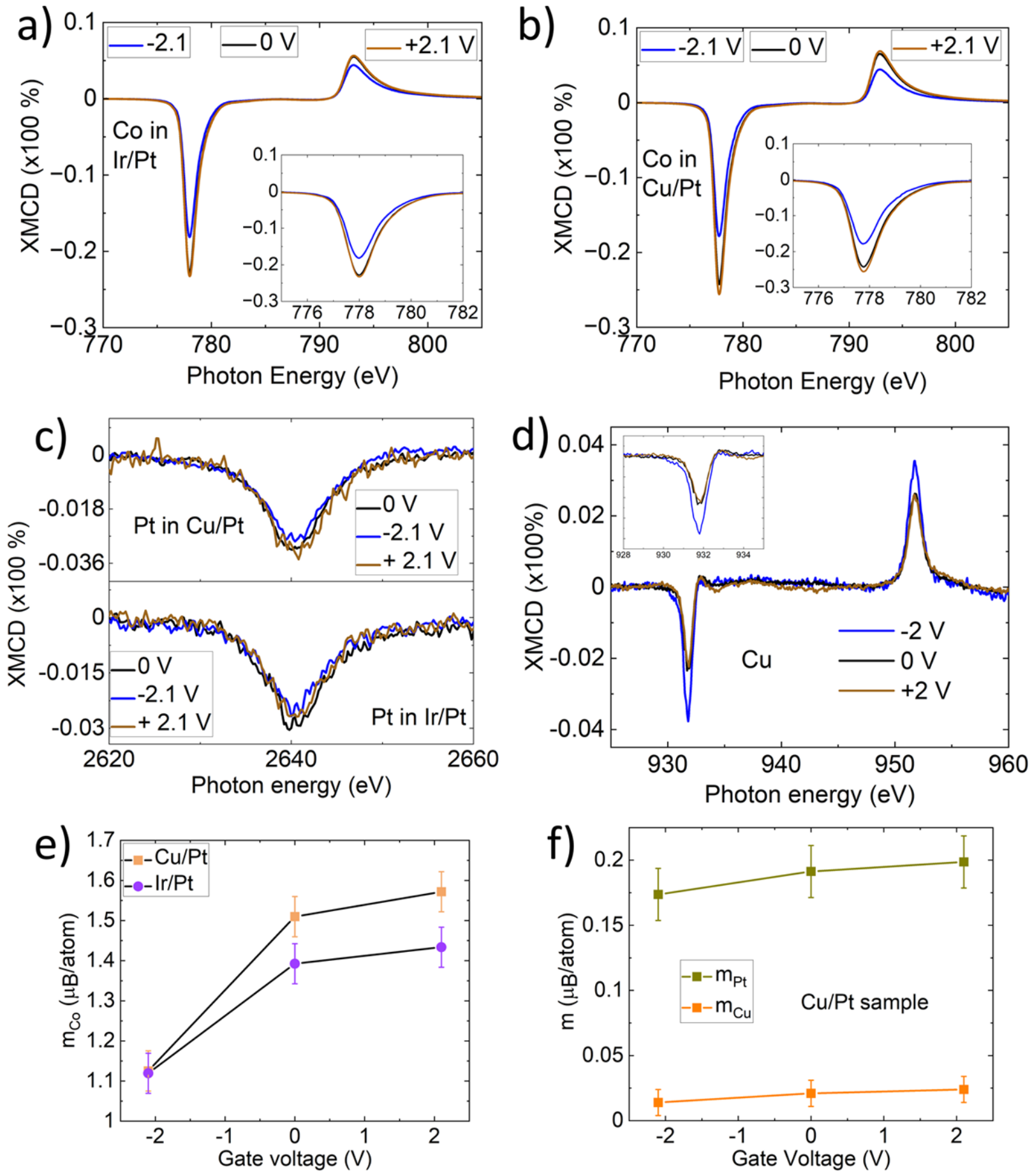

**Figure 3.** XMCD spectra of (a) Ir/Pt and (b) Cu/Pt. Both samples show low dichroism under a negative voltage. However, under a positive voltage, they exhibit higher dichroism compared to the as-grown sample. (c) Pt and (d) Cu show low dichroism. e) Co moment per atom for the Cu/Pt and Ir/Pt samples. The magnetic moments are decreasing with negative voltages but within error bars for the positive voltages. At −2 V the magnetic moments are close; the reason might be due to the high oxidation rate of Cu. f) Polarization of Pt and Cu in Cu/Pt samples. The polarization of Cu is significantly low compared to that of Pt.

contrast, the difference between the positive voltage and those of the as-grown samples is negligible. The oxygen ion transport mechanism is the same for both samples, but it is less pronounced in the Cu/Pt sample. The Cu-$L_3$ XAS peak at approximately 932 eV corresponds to Cu metal,[37,38] while the other peaks are attributed to copper-oxide phases (Figure 2(e)). Under a negative voltage, the metallic peak decreases, and the Cu-oxide phase peak (at approximately 934 eV) becomes significantly larger. In contrast, the positive voltage causes no noticeable change, consistent with the behavior observed in the Co-L and Pt-M data. We were unable to measure Ir due to its very low scattering cross-section (the XAS signal at the Ir-M absorption edge is presented in the Supporting Information, Figure S8) at the Ir-$M_3$ edge, making it less sensitive to the applied voltages. Overall, the XAS data collected at the Co, Cu-L, and Pt-M edges demonstrate controlled oxygen ion transport under the applied voltages. As discussed earlier (Figure 1(b)), the applied voltages induce oxygen migration, changing the magnetization of the samples. Here, we analyze the element-specific XMCD spectra and discuss the derived magnetic moments.

In order to understand the influence of changing the oxidation at the CoB and adjacent layers on the magnetic properties, we also measured the XMCD spectra. Figure 3 shows the Co- L XMCD spectra for the Ir/Pt samples (a) and Cu/Pt samples (b). In both systems, a positive voltage results in slightly larger dichroism (as indicated by the Co-L3 XMCD peak intensity in Figure 3(a, b)), suggesting increased magnetization.[39] In contrast, the negative voltage produces the lowest dichroism. This trend is consistently observed in both systems. Additionally, Pt and Cu atoms exhibit nonzero dichroism (Figure 3c,d,f), confirming the existence of a magnetic proximity effect in the samples, although the changes with applied voltages may not be significant. Using the XMCD sum rules,[40] we calculated the spin ($m_s$) and orbital ($m_l$) magnetic moments for individual elements. Figure 3(e) presents the total magnetic moment ($m_{tot} = m_l + m_s$) of Co atoms in the Cu/Pt and Ir/Pt systems as a function of the applied voltage. The moment decreases significantly where a negative voltage was applied, which is consistent with the magnetometry data. Samples where a positive voltage was applied do not show as significant an increase in moment as the magnetometry data suggests. The magnetometry data, however, was taken immediately after applying the voltage, while the synchrotron measurements were some weeks afterward for logistical reasons. This difference is consistent with our interpretation that the positive voltage moves oxygen slightly away from the CoB interface; however, it is still free to migrate back after the voltage is removed. In contrast, the negative voltage drives the oxygen toward the interface where it bonds to the Co and remains. Figure S9 in the Supporting Information demonstrates this with a series of magnetic hysteresis loops measured over the course of 15 weeks for negatively gated and positively gated samples. Dugato et al. (2023)[41] examined the Co thickness-dependent magnetic moment using the XMCD technique. They observed that the magnetic moment ($Co_m$) increases with Co thickness (5−15 Å), ranging from 1.2 $\mu_B$ to 1.7 $\mu_B$, and observed that the $Co_m$ is lower for a thickness of 5 Å. Figure 3(f) displays the $m_s$ values for Pt and Cu in the Cu/Pt sample. The Pt moments at the $M_3$ absorption edge were calculated using the protocols outlined by Valvidares et al. (2016)[42] and Vasili et al. (2018).[36] Lau et al. (2019)[43] investigated Pt/Co/Pt and Pt/Co/Ir systems using XMCD, determining the total magnetic moment of Pt ($Pt_m$) to range from 0.14 $\mu_B$/hole to 0.26 $\mu_B$/hole. Despite our Pt layer being only 6 Å thick, the observed Pt moments are comparable to these reported values. We also applied the XMCD sum rules to the Cu-$L_{2,3}$ edges, finding that our data align well with previously reported Co/Cu/Pt systems.[38] Parreiras et al. (2021) studied the magnetic moment of Co/$Cu_3Au$(100),[44] calculating the Co moment to be approximately ≈1.75 $\mu_B$ and the Cu moment to be ≈0.024 $\mu_B$. Therefore, the calculated moments fall within the range of the literature values. The changes in Cu-L or Pt-M moments with applied voltage are minimal and are within the error bars.

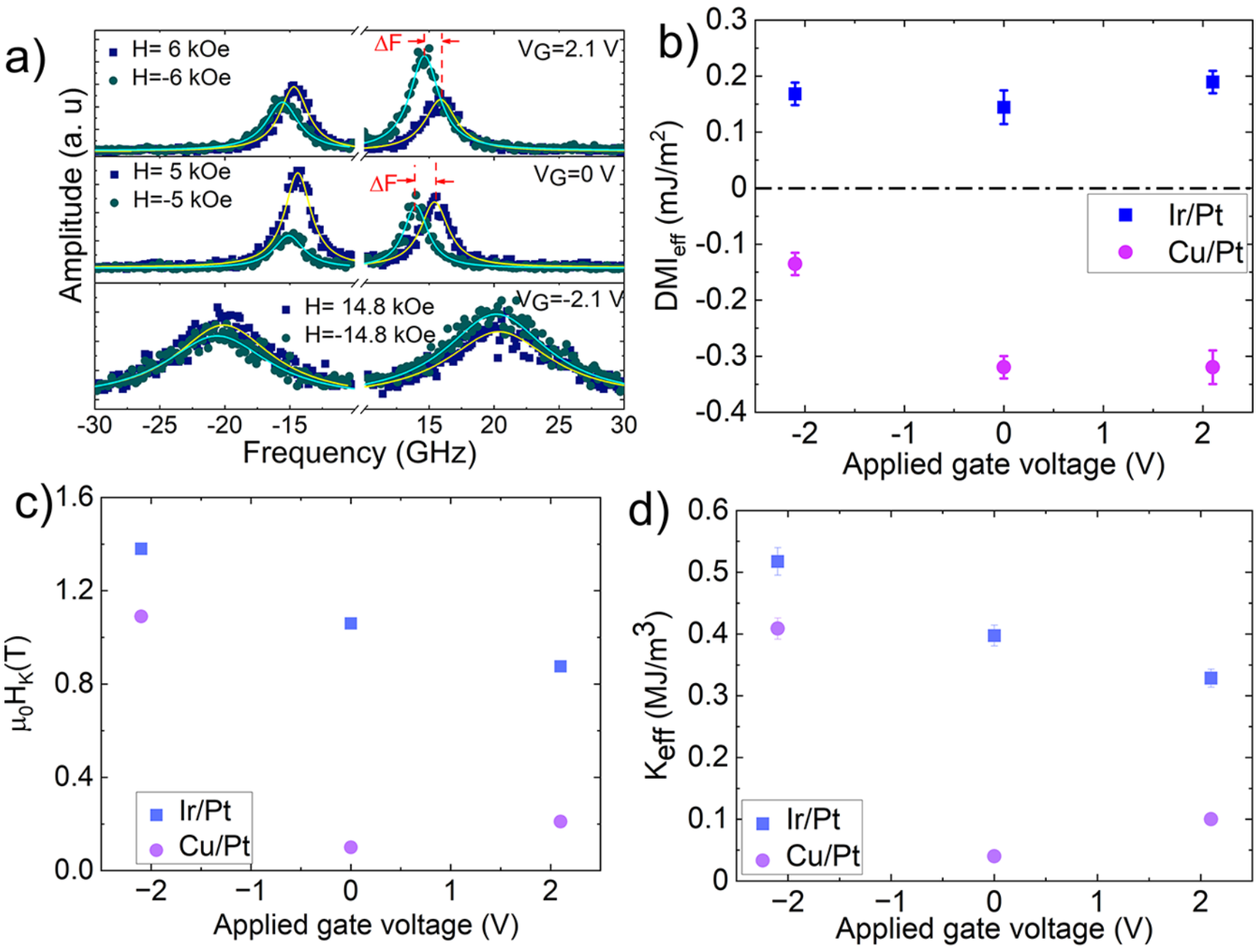

**Figure 4.** (a) Typical BLS spectra showing the frequency difference between the Stokes and anti-Stokes peaks of the Cu/Pt sample, used to calculate the $DMI_{eff}$. (b) $DMI_{eff}$ values of the Ir/Pt and Cu/Pt samples. The $DMI_{eff}$ of the Cu/Pt sample changes significantly upon the application of a negative voltage. (c) $H_K$ and (d) $K_{eff}$ of Ir/Pt and Cu/Pt samples: The application of a negative voltage increases both the anisotropy field and the effective anisotropy constant, whereas the application of a positive voltage results in a decrease in these parameters in Ir/Pt samples and a small increase in Cu/Pt.

## Interfacial Magnetic Properties: Effective DMI and Effective Anisotropy

Having established the structural and chemical modifications at the upper CoB interface, we now turn our attention to the resultant changes in magnetic properties. The Brillouin light scattering (BLS) spectrometry technique exploits the nonreciprocal propagation of spin waves (SWs) induced $DMI_{eff}$ in magnetic films. Consequently, the $DMI_{eff}$ can be calculated from the frequency mismatch ΔF between the Stokes and anti-Stokes frequency lines obtained from the measurements (Figure 4(a)). The $DMI_{eff}$ is calculated using the equation in Supporting Information Section 4, which is shown in Figure 4(b). The Pt layer at the bottom induces a negative $DMI_{eff}$, whereas the top Pt layer induces a positive $DMI_{eff}$. The Ir/Pt sample shows a positive $DMI_{eff}$, suggesting a left-handed (counterclockwise) chirality.[45] The Ir layer induces its own $DMI_{eff}$, and its sign is the same as that of the top Pt layer, resulting in an overall positive $DMI_{eff}$. Ir/Pt samples exhibit a lower $DMI_{eff}$ compared with Cu/Pt samples. In the Cu/Pt case, the Cu layer screens the contribution of the top Pt layer to $DMI_{eff}$, making the system more asymmetric and therefore leading to a larger negative $DMI_{eff}$. In contrast, as the Ir layer contributes to the $DMI_{eff}$, the DMI signs of Pt/CoB and CoB/Ir are opposite, and the total DMI becomes relatively weak. With applied negative voltage, the $DMI_{eff}$ of the Cu/Pt sample decreases significantly. With applied voltages, the Ir/Pt samples show a similar $DMI_{eff}$. Chen et al. (2015) also reported positive $DMI_{eff}$ for Co/Ir samples.[46] Diez et al. (2019)[11] studied the magneto-ionic effect on DMI in a Pt/Co/$HfO_2$ structure. They observed a reduction in DMI with oxygen ion migration, and the effect was not reversible. Under a negative voltage, oxygen traveled to the Pt/Co interface and decoupled it, thus reducing DMI. We have also observed a reduction in $DMI_{eff}$ in Cu/Pt samples with a negative voltage. The reason is the effect of the top Pt layer, where the oxygen decouples the Cu atoms. Therefore, Pt interacts with CoB, resulting in a decrease in total $DMI_{eff}$. The applied voltage changes the upper interface, changing the strength of DMI.[47−49] On the other hand, since Ir and Pt generate the same $DMI_{eff}$, the decoupling of Ir does not affect the total $DMI_{eff}$. These findings also indicate that with a negative voltage the migration of oxygen does not reach the bottom Pt/CoB layer for both samples.

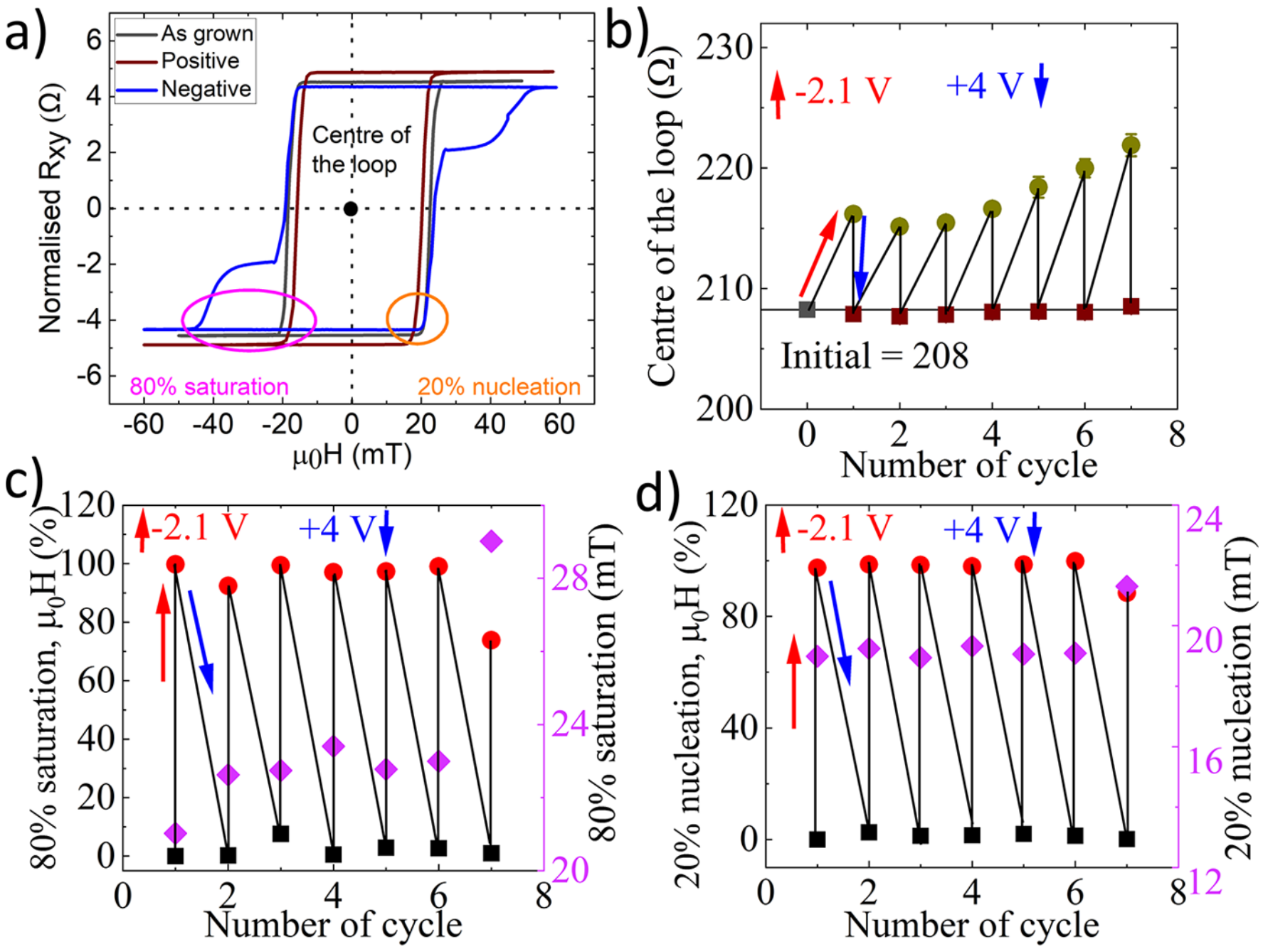

**Figure 5.** (a) The hysteresis loops are shown for as-grown, negatively gated, and positively gated samples. The electrical contacts for the AHE measurements were placed at the edge of the thin film, resulting in contributions from both gated and nongated regions. Reversibility of the Ir/Pt sample as a function of cycle number: b) center of the loop after applied voltage −2.1 V for 120 s, where +4 V is used to reverse the effect. c) Reversibility at 80% of saturation. d) Reversibility at 20% nucleation. The sample demonstrates almost full reversibility with the slight discrepancy attributed to the movement of the top (ITO glass) electrode.

Figure 4(c-d) shows the measured anisotropy field and the effective anisotropy ($K_{eff}$), calculated from the equation in the Supporting Information, Section 4 and Figure S5. The ungated and positively gated Ir/Pt samples exhibit significantly higher effective anisotropy compared to that of the equivalent Cu/Pt samples. This is consistent with the work of Benguettat et al. (2020),[48] who studied the anisotropy of the Co/Ir and Co/Cu interfaces, observing a lower anisotropy in the Co/Cu samples, originating mainly from the bottom Pt−Co interface. In contrast, a strong interlayer coupling of Ir contributes to an increase in anisotropy.[50,51] In the present work, however, the Cu and Ir are incomplete "dusting" layers and the Co is not a well ordered hcp or fcc surface but present in a highly disordered CoB layer, and so the anisotropy energies are quantitatively comparable. With the negatively gated samples, both Ir/Pt and Cu/Pt show an increase in anisotropy field and effective anisotropies as both the dusting layers (Ir or Cu) and Co in the CoB layer oxidize, with the increase in the interfacial anisotropy being due to the hybridization of Co-3d and O-2p.[52−56]

## Reversibility Investigation Using Anomalous Hall Effect (AHE)

As previously observed in the studied multilayer systems, the negative voltage has a larger effect than the positive voltage. Thus, a positive voltage was applied to test the reversibility of the effects of the negative gate voltage. The negative voltage increases the AH resistance, while the positive voltage decreases it. Figure 5(a) shows the AHE hysteresis loop, which was taken outside the gated area (Figure 1(a)). The shape of the hysteresis curve changes under different gate voltages as a result of the E-field-induced relocation of oxygen ions within the gated area. As a result, the hysteresis loop with applied voltage is a mixture of gated (G) and ungated (U) regions. It is assumed that the data can be decomposed into components that are symmetric and antisymmetric with respect to the magnetic field. Thus, the antisymmetric part, identified as $R_{xy}$, is considered to be the Hall signal. The symmetric part, which remains constant with the magnetic field, is denoted as $R_{xx}$ (see Supporting Information, Figure S7). The changes observed in Figure 5(a) are attributed to the change in anisotropy caused by the negative voltage. Under a negative gate voltage, the coercive field in the gated region increases; therefore, higher-field switching arises from the gating effect. In contrast, under a positive field, domain switching in the gated region begins earlier, leading to a reduced switching field for the entire sample. To quantify these effects, we used a convenient metric: the magnetic field at which the device reaches 80% saturation, representing the state just before full saturation, and the field at which 20% of the gated area has switched, indicating the initial stage of domain nucleation and motion.

Figure 5(b) illustrates the center of the hysteresis loop at the initial position, where an applied −2.1 V for 120 s sequentially increases resistance in each cycle. In cycle 1, the system was reversed with an applied +4 V for 60 s, while a longer exposure time was required for the following cycles. Figures 5(c) and 5(d) show reversibility at 80% saturation and 20% nucleation, respectively. The left axis represents the reversibility with respect to the previous cycle, whereas the right axis indicates the magnetic field value for each cycle. In both cases, the system is almost entirely reversible. Reversibility cycles are shown in the Supporting Information, Figure S6. The ITO electrode is placed on liquid using conductive silver paste; as a result, it may shift slightly on the sample. Due to the movement of the ITO electrode, the seventh cycle exhibits a large deviation, approximately 29 mT at 80% saturation.

## CONCLUSIONS

In this work, we have presented the direct observation of oxygen migration under an electric field and its impact on magnetic properties. The controlled migration of oxygen impacts the sample structure and alters the chemical state of the interface, leading to the formation of CoO, which is the key to the nonvolatile yet reversible modulation of the magnetic properties. Measurements consistently show CoO formation as a result of negative gate voltage, increasing anisotropy from $0.04 \pm 0.01$ MJ/m$^3$ to $0.4 \pm 0.02$ MJ/m$^3$ for Cu/Pt samples and from $0.4 \pm 0.02$ MJ/m$^3$ to $0.51 \pm 0.02$ MJ/m$^3$ for Ir/Pt samples. On the other hand, it decreases the $DMI_{eff}$ of Cu/Pt from $0.32 \pm 0.01$ mJ/m$^2$ to $0.14 \pm 0.01$ mJ/m$^2$. The magnetic moment also decreases from 1.40 $\mu_B$ to 1.11 $\mu_B$ for the Ir/Pt samples and from 1.5 $\mu_B$ to 1.11 $\mu_B$ for the Cu/Pt samples. The reversibility of the magnetoionic effect in the thin film heterostructure is useful for designing field-programmable spintronic devices.

## EXPERIMENTAL SECTION

These magnetic multilayers are grown on $Si/SiO_2$ (300 nm) substrates by using DC magnetron sputtering at room temperature in the Royce deposition system at the University of Leeds, UK. Prior to deposition, the substrates went through a cleaning process in an ultrasonic water bath using acetone, isopropanol, and deionized (DI) water. Subsequently, they were baked at 110 °C for 1 min and then subjected to plasma ashing for an additional minute to remove organic contaminants. The oxygen plasma contains oxygen ions, radicals, and ozone. These species react with the organic contaminants and produce gases. Our sputtering system operates within an ultrahigh vacuum environment with a base pressure of $10^{-9}$ mbar and a growth pressure of $10^{-3}$ mbar. To achieve uniformity in the films, the substrate is rotated at 120°/s during the deposition process. Ta was grown at 80 W, CoB was grown at 50 W, and Ir, Pt, and Cu were grown at 18 W. A 25 Å $HfO_2$ was grown for the application of the E-field by ex situ thermal ALD (90 °C). The precursor was tetrakis(dimethylamido)hafnium $Hf(NMe_2)_4$. The precursor gas is introduced into the chamber, where it chemisorbs onto the substrate (a metallic film in this case), forming a monolayer of Hf. Nitrogen gas is then used to purge any excess precursor materials. Water vapor is subsequently introduced to oxidize the deposited Hf, resulting in the formation of $HfO_2$.

Most of the measurements were performed exclusively on the gated area; therefore, the subsequent measurement was destructive or involved removing the gate electron, and so repeated reverse application was not feasible. However, in the reversibility study, the same samples were first exposed to −2.1 V, followed by +2.1 V to reverse the effect. This process was repeated over several cycles. During AHE, both the gated and the ungated regions were measured. The AHE contacts were positioned approximately 1 mm away from the gated region. The ITO glass electrode is placed on the liquid and connected using silver paste, which is allowed to dry for approximately 5−10 min. Because one end of the electrode is fixed by the silver paste, while the ITO rests on the liquid surface, the setup is vibration-sensitive and the ITO can shift slightly relative to the original place.

All data presented here were obtained from 6 Å of the top layer of Pt, except the reversibility data (Figure 5), which were taken from 4 Å of the top layer of Pt. Additionally, Figure 1c shows the variation in the CoB thickness.

The BLS measurements were performed at the LSPM, France. The measurement for DMI is taken to a maximum wave vector $k_{sw} = 20.45$ $\mu m^{-1}$, which corresponds to the incident angle of 60° for positive and negative magnetic field.

The anisotropy is measured by the angular dependence of the anomalous Hall effect (AHE). The sample is placed in a constant out-of-plane field of 500 mT and rotated from 0° to 360°. The resulting Hall voltage was recorded as a function of angle. The parabolic part of the data is fitted with a parabola function, which gives the anisotropy field.[57]

X-ray absorption spectroscopy and magnetic circular dichroism (XAS and XMCD) measurements were carried out at the BL-29 BOREAS beamline of the ALBA synchrotron, Spain.[58] The XMCD spectra were obtained by taking the difference between the XAS of left (-) and right (+) circularly polarized light. In the XMCD measurements, magnetic fields up to ±6 T were applied either in grazing or normal incidence to the film plane. The spectra were measured in total electron yield (TEY) mode. The spectra were normalized to the incoming flux, as measured by the drain current signal on a gold mesh upstream of the sample.

## ASSOCIATED CONTENT

### Data Availability Statement

Zenodo: 10.5281/zenodo.18865446

### Supporting Information

The Supporting Information is available free of charge at https://pubs.acs.org/doi/10.1021/acsaelm.6c00200.

Section 1: Kerr microscopy and XMCD hysteresis loops; Section 2: Toy model; Section 3: XPS analysis; Section 4: DMI and anisotropy calculations; Section 5: Reversibility cycles and $R_{xx}$; Section 6: XAS signals of Ir and Pt; and Section 7: Nonvolatility effect. (PDF)

## AUTHOR INFORMATION

### Corresponding Authors

**Md Golam Hafiz** − *School of Physics and Astronomy, University of Leeds, Leeds LS2 9JT, U.K.; Laboratoire des Sciences des Procédés et des Matériaux, 93430 Villetaneuse, France*; orcid.org/0009-0008-1787-1212; Email: hafizmdgolam927@gmail.com

**Gavin Burnell** − *School of Physics and Astronomy, University of Leeds, Leeds LS2 9JT, U.K.*; Email: g.burnell@leeds.ac.uk

### Authors

**Hari Babu Vasili** − *School of Physics and Astronomy, University of Leeds, Leeds LS2 9JT, U.K.*

**Philippa Shepley** − *School of Physics and Astronomy, University of Leeds, Leeds LS2 9JT, U.K.*

**Mannan Ali** − *School of Physics and Astronomy, University of Leeds, Leeds LS2 9JT, U.K.*

**Andrew J. Britton** − *School of Chemical and Process Engineering, University of Leeds, Leeds LS2 9JT, U.K.*

**Oscar Cespedes** − *School of Physics and Astronomy, University of Leeds, Leeds LS2 9JT, U.K.*; orcid.org/0000-0002-5249-9523

**Weibin Li** − *ALBA Synchrotron Light Source, E-08290 Barcelona, Catalonia, Spain*; orcid.org/0009-0003-4037-113X

**Manuel Valvidares** − *ALBA Synchrotron Light Source, E-08290 Barcelona, Catalonia, Spain*; orcid.org/0000-0003-4895-8114

**Rohit Pachat** − *Centre de Nanosciences et de Nanotechnologies, CNRS, Université Paris-Saclay, 91120 Palaiseau, France*

**Wilfried Fotso** − *Laboratoire des Sciences des Procédés et des Matériaux, 93430 Villetaneuse, France*

**Mourad Cherif** − *Laboratoire des Sciences des Procédés et des Matériaux, 93430 Villetaneuse, France;* orcid.org/0000-0003-4350-9379

**Shimpei Ono** − *International Center for Synchrotron Radiation Innovation Smart (SRIS), Sendai 980-8572, Japan*

**Yves Roussigne** − *Laboratoire des Sciences des Procédés et des Matériaux, 93430 Villetaneuse, France;* orcid.org/0000-0001-7698-8092

**Mohamed Belmeguenai** − *Laboratoire des Sciences des Procédés et des Matériaux, 93430 Villetaneuse, France*

Complete contact information is available at:
https://pubs.acs.org/10.1021/acsaelm.6c00200

## Notes

The authors declare no competing financial interest.

## ■ ACKNOWLEDGMENTS

We acknowledge the funding support from the European Union's Horizon 2020 research and innovation program under Marie Sklodowska-Curie Grants No. 860060'Magnetism and the effect of Electric Field'(MagnEFi). The authors also acknowledge A. Huxtable, M. Rogers, and C. Knox for assistance in developing experimental methods. We acknowledge support from the Henry Royce Institute (EPSRC grants: EP/P022464/1, EP/R00661*X*/1), which funded the VXSF facilities (https://engineering.leeds.ac.uk/vxsf) and the Royce deposition system within the Bragg Center for Materials Research at Leeds. We also acknowledge the funding of French National Research Agency (ANR) within the ANR program EHIS (ANR-21-CE42-0003). We acknowledge XAS, XMCD experiments at BOREAS beamline (ALBA synchrotron) through proposal No. 2023097865.

## ■ REFERENCES

(1) Bauer, U.; Yao, L.; Tan, A. J.; Agrawal, P.; Emori, S.; Tuller, H. L.; Van Dijken, S.; Beach, G. S. Magneto-ionic control of interfacial magnetism. *Nat. Mater.* **2015**, *14*, 174−181.

(2) Pachat, R.; Ourdani, D.; van der Jagt, J. W.; Syskaki, M. A.; Di Pietro, A.; Roussigne, Y.; Ono, S.; Gabor, M. S.; Cherif, M.; Durin, G.; Langer, J.; Belmeguenai, M.; Ravelosona, D.; Diez, L. H. Multiple magneto-ionic regimes in Ta/Co20Fe60B20/HfO2. *Physical Review Applied* **2021**, *15*, No. 064055.

(3) Tan, A. J.; Huang, M.; Avci, C. O.; Büttner, F.; Mann, M.; Hu, W.; Mazzoli, C.; Wilkins, S.; Tuller, H. L.; Beach, G. S. Magneto-ionic control of magnetism using a solid-state proton pump. *Nat. Mater.* **2019**, *18*, 35.

(4) Bauer, U.; Emori, S.; Beach, G. S. Voltage-controlled domain wall traps in ferromagnetic nanowires. *Nat. Nanotechnol.* **2013**, *8*, 411−416.

(5) Obinata, A.; Hibino, Y.; Hayakawa, D.; Koyama, T.; Miwa, K.; Ono, S.; Chiba, D. Electric-field control of magnetic moment in Pd. *Sci. Rep.* **2015**, *5*, 14303.

(6) Radaelli, G.; Petti, D.; Plekhanov, E.; Fina, I.; Torelli, P.; Salles, B. R.; Cantoni, M.; Rinaldi, C.; Gutiérrez, D.; Panaccione, G.; Varela, M.; Picozzi, S.; Fontcuberta, J.; Bertacco, R. Electric control of magnetism at the Fe/BaTiO3 interface. *Nat. Commun.* **2014**, *5*, 3404.

(7) Liu, Y. T.; Ono, S.; Agnus, G.; Adam, J. P.; Jaiswal, S.; Langer, J.; Ocker, B.; Ravelosona, D.; Herrera Diez, L. Electric field controlled domain wall dynamics and magnetic easy axis switching in liquid gated CoFeB/MgO films. *J. Appl. Phys.* **2017**, *122*, No. 133907.

(8) Murray, P. D.; Jensen, C. J.; Quintana, A.; Zhang, J.; Zhang, X.; Grutter, A. J.; Kirby, B. J.; Liu, K. Electrically Enhanced Exchange Bias via Solid-State Magneto-ionics. *ACS Appl. Mater. Interfaces* **2021**, *13*, 38916−38922.

(9) Jensen, C. J.; Quintana, A.; Quarterman, P.; Grutter, A. J.; Balakrishnan, P. P.; Zhang, H.; Davydov, A. V.; Zhang, X.; Liu, K. Nitrogen-Based Magneto-ionic Manipulation of Exchange Bias in CoFe/MnN Heterostructures. *ACS Nano* **2023**, *17*, 6745−6753.

(10) Zehner, J.; Wolf, D.; Hasan, M. U.; Huang, M.; Bono, D.; Nielsch, K.; Leistner, K.; Beach, G. S. Magnetoionic control of perpendicular exchange bias. *Physical Review Materials* **2021**, *5*, L061401.

(11) Herrera Diez, L.; et al. Nonvolatile Ionic Modification of the Dzyaloshinskii-Moriya Interaction. *Physical Review Applied* **2019**, *12*, No. 034005.

(12) Zehner, J.; Soldatov, I.; Schneider, S.; Heller, R.; Khojasteh, N. B.; Schiemenz, S.; Fähler, S.; Nielsch, K.; Schäfer, R.; Leistner, K. Voltage-Controlled Deblocking of Magnetization Reversal in Thin Films by Tunable Domain Wall Interactions and Pinning Sites. *Advanced Electronic Materials* **2020**, *6*, 2000406.

(13) Ameziane, M.; Huhtasalo, J.; Flajšman, L.; Mansell, R.; van Dijken, S. Solid-State Lithium Ion Supercapacitor for Voltage Control of Skyrmions. *Nano Lett.* **2023**, *23*, 3167−3173.

(14) Shepley, P. M.; Tunnicliffe, H.; Shahbazi, K.; Burnell, G.; Moore, T. A. Magnetic properties, domain-wall creep motion, and the Dzyaloshinskii-Moriya interaction in Pt/Co/Ir thin films. *Phys. Rev. B* **2018**, *97*, No. 134417.

(15) Zeissler, K.; Finizio, S.; Barton, C.; Huxtable, A. J.; Massey, J.; Raabe, J.; Sadovnikov, A. V.; Nikitov, S. A.; Brearton, R.; Hesjedal, T.; van der Laan, G.; Rosamond, M. C.; Linfield, E. H.; Burnell, G.; Marrows, C. H. Diameter-independent skyrmion Hall angle observed in chiral magnetic multilayers. *Nat. Commun.* **2020**, *11*, 428.

(16) Büttner, F.; et al. Dynamics and inertia of skyrmionic spin structures. *Nat. Phys.* **2015**, *11*, 225−228.

(17) Fert, A.; Levy, P. M. Role of anisotropic exchange interactions in determining the properties of spin-glasses. *Phys. Rev. Lett.* **1980**, *44*, 1538.

(18) Hrabec, A.; Porter, N. A.; Wells, A.; Benitez, M. J.; Burnell, G.; McVitie, S.; McGrouther, D.; Moore, T. A.; Marrows, C. H. Measuring and tailoring the Dzyaloshinskii-Moriya interaction in perpendicularly magnetized thin films. *Physical Review B - Condensed Matter and Materials Physics* **2014**, *90*, 020402.

(19) He, M.; Xu, T.; Gao, Y.; Hu, C.; Cai, J.; Zhang, Y. Mixed-Type Skyrmions in Symmetric Pt/Co/Pt Multilayers at Room Temperature. *Materials* **2022**, *15*, 8272.

(20) Nath, J.; et al. Mechanism of Spin Orbit Torques in Platinum Oxide Systems. *Adv. Electron. Mater.* **2022**, *8*, 2101335.

(21) Xie, H.; Yuan, J.; Luo, Z. In-situ study of oxygen exposure effect on spin-orbit torque in Pt/Co bilayers in ultrahigh vacuum. *Sci. Rep.* **2019**, *9*, 17254.

(22) Darwin, E.; Tomasello, R.; Shepley, P. M.; Satchell, N.; Carpentieri, M.; Finocchio, G.; Hickey, B. J.; et al. Antiferromagnetic interlayer exchange coupled Co68B32/Ir/Pt multilayers. *Sci. Rep.* **2024**, *14*, 95.

(23) Lavrijsen, R.; Verheijen, M. A.; Barcones Campo, B.; Kohlhepp, J. T.; Swagten, H. J. M.; Koopmans, B. Enhanced field-driven domain wall motion in Pt/Co/Pt and Pt/$Co_{68}B_{32}$/Pt strips. *Appl. Phys. Lett.* **2011**, *98*, No. 132502.

(24) Tan, H. K.; Lim, R. J.; Seng, H. L.; Shanmugam, J.; Ko, H. Y.; Cheng, X. M.; Putra, V.; Xing, Z. X.; Soumyanarayanan, A.; Ho, P. Intermixing induced anisotropy variations in CoB-based chiral multilayer films. *J. Phys. D: Appl. Phys.* **2021**, *54*, No. 354003.

(25) Jian, L.; Choi, G.-M. Using Amorphous CoB Alloy as Transducer to Detect Acoustic Propagation and Heat Transport at Interface. *Applied Sciences* **2021**, *11*, 5155.

(26) Buttner, F.; Moutafis, C.; Bisig, A.; Wohlhuter, P.; Gunther, C. M.; Mohanty, J.; Geilhufe, J.; Schneider, M.; Schmising, C.v.K.; Schaffert, S.; et al. Magnetic states in low-pinning high-anisotropy material nanostructures suitable for dynamic imaging. *Phys. Rev. B* **2013**, *87*, No. 134422.

(27) Ono, S.; Seki, S.; Hirahara, R.; Tominari, Y.; Takeya, J. High-mobility, low-power, and fast-switching organic field-effect transistors with ionic liquids. *Appl. Phys. Lett.* **2008**, *92*, No. 103313.

(28) Leighton, C. Electrolyte-based ionic control of functional oxides. *Nat. Mater.* **2019**, *18*, 13−18.
(29) Pachat, R.; et al. Magneto-Ionics in Annealed W/CoFeB/HfO2 Thin Films. *Advanced Materials Interfaces* **2022**, *9*, 2200690.
(30) Gu, Y.; Song, C.; Wang, Q.; Hu, W.; Liu, W.; Pan, F.; Zhang, Z. Emerging opportunities for voltage-driven magneto-ionic control in ferroic heterostructures. 2021.
(31) Nichterwitz, M.; Honnali, S.; Kutuzau, M.; Guo, S.; Zehner, J.; Nielsch, K.; Leistner, K. Advances in magneto-ionic materials and perspectives for their application. *APL Materials* **2021**, *9*, No. 030903.
(32) Balan, C.; Peña Garcia, J.; Fassatoui, A.; Vogel, J.; Chaves, D. D. S.; Bonfim, M.; Rueff, J. P.; Ranno, L.; Pizzini, S. Tuning the Dynamics of Chiral Domain Walls of Ferrimagnetic Films by Magnetoionic Effects. *Physical Review Applied* **2022**, *18*, No. 034065.
(33) Li, H. B.; et al. Electric-field control of ferromagnetism through oxygen ion gating. *Nat. Commun.* **2017**, *8*, 2156.
(34) Satchell, N.; Mitchell, T.; Shepley, P. M.; Darwin, E.; Hickey, B. J.; Burnell, G. Pt and CoB trilayer Josephson $\pi$ junctions with perpendicular magnetic anisotropy. *Sci. Rep.* **2021**, *11*, 11173.
(35) Ma, Z.; Fuentes-Rodriguez, L.; Tan, Z.; Pellicer, E.; Abad, L.; Herrero-Martín, J.; Menéndez, E.; Casañ-Pastor, N.; Sort, J. Wireless magneto-ionics: voltage control of magnetism by bipolar electro-chemistry. *Nat. Commun.* **2023**, *14*, 6486.
(36) Vasili, H. B.; Gamino, M.; Gàzquez, J.; Sánchez, F.; Valvidares, M.; Gargiani, P.; Pellegrin, E.; Fontcuberta, J. Magnetoresistance in Hybrid Pt/CoFe2O4 Bilayers Controlled by Competing Spin Accumulation and Interfacial Chemical Reconstruction. *ACS Appl. Mater. Interfaces* **2018**, *10*, 12031−12041.
(37) Samant, M. G.; Stöhr, J.; Parkin, S. S.; Held, G. A.; Hermsmeier, B. D.; Herman, F.; Van Schilfgaarde, M.; Duda, L. C.; Mancini, D. C.; Wassdahl, N.; Nakajima, R. Induced spin polarization in Cu spacer layers in Co/Cu multilayers. *Phys. Rev. Lett.* **1994**, 72.1112
(38) Okabayashi, J.; Koyama, T.; Suzuki, M.; Tsujikawa, M.; Shirai, M.; Chiba, D. Induced perpendicular magnetization in a Cu layer inserted between Co and Pt layers revealed by x-ray magnetic circular dichroism. *Sci. Rep.* **2017**, *7*, 46132.
(39) Gilbert, D. A.; Grutter, A. J.; Arenholz, E.; Liu, K.; Kirby, B. J.; Borchers, J. A.; Maranville, B. B. Structural and magnetic depth profiles of magneto-ionic heterostructures beyond the interface limit. *Nat. Commun.* **2016**, *7*, 12264.
(40) Chen, C. T.; Idzerda, Y. U.; Lin, H. J.; Smith, N. V.; Meigs, G.; Chaban, E.; Ho, G. H.; Pellegrin, E.; Sette, F. Experimental confirmation of the x-ray magnetic circular dichroism sum rules for iron and cobalt. *Phys. Rev. Lett.* **1995**, *75*, 152.
(41) Dugato, D. A.; Brandão, J.; Béron, F.; da Silva, R. B.; Flewett, S.; Shapiro, D. A.; Cezar, J. C.; Dorneles, L. S.; Mori, T. J. Proximity induced moment at Pt/Co interfaces and isolated skyrmion bubble stabilization at zero magnetic field. *J. Magn. Magn. Mater.* **2023**, *566*, No. 170305.
(42) Valvidares, M.; Dix, N.; Isasa, M.; Ollefs, K.; Wilhelm, F.; Rogalev, A.; Sánchez, F.; Pellegrin, E.; Bedoya-Pinto, A.; Gargiani, P.; Hueso, L. E.; Casanova, F.; Fontcuberta, J. Absence of magnetic proximity effects in magnetoresistive Pt/CoF e2 O4 hybrid interfaces. *Phys. Rev. B* **2016**, *93*, No. 214415.
(43) Lau, Y. C.; Chi, Z.; Taniguchi, T.; Kawaguchi, M.; Shibata, G.; Kawamura, N.; Suzuki, M.; Fukami, S.; Fujimori, A.; Ohno, H.; Hayashi, M. Giant perpendicular magnetic anisotropy in Ir/Co/Pt multilayers. *Physical Review Materials* **2019**, *3*, No. 104419.
(44) Parreiras, S. O.; Cabral, L. A.; Lourenço, R. V.; Cotta, A. A.; Schio, P.; Cezar, J. C.; Gastelois, P. L.; da Silva, E. Z.; Macedo, W. A. Induced magnetization in Cu atoms at the Fe-Co/Cu3Au(001) interface: X-ray magnetic circular dichroism experiments and theoretical results. *Appl. Surf. Sci.* **2021**, *548*, No. 149215.
(45) Chen, R.; Wang, X.; Cheng, H.; Lee, K. J.; Xiong, D.; Kim, J. Y.; Li, S.; Yang, H.; Zhang, H.; Cao, K.; Kläui, M.; Peng, S.; Zhang, X.; Zhao, W. Large Dzyaloshinskii-Moriya interaction and room-temper-ature nanoscale skyrmions in CoFeB/MgO heterostructures. *Cell Reports Physical Science* **2021**, *2*, No. 100618.
(46) Chen, G.; N'Diaye, A. T.; Wu, Y.; Schmid, A. K. Ternary superlattice boosting interface-stabilized magnetic chirality. *Appl. Phys. Lett.* **2015**, *106*, No. 062402.
(47) Belmeguenai, M.; Roussigné, Y.; Chérif, S. M.; Stashkevich, A.; Petrisor, T.; Nasui, M.; Gabor, M. S. Influence of the capping layer material on the interfacial Dzyaloshinskii-Moriya interaction in Pt/Co/capping layer structures probed by Brillouin light scattering. *J. Phys. D: Appl. Phys.* **2019**, *52*, No. 125002.
(48) Benguettat-El Mokhtari, I.; Ourdani, D.; Roussigné, Y.; Mos, R. B.; Nasui, M.; Kail, F.; Chahed, L.; Chérif, S. M.; Stashkevich, A.; Gabor, M.; Belmeguenai, M. Perpendicular magnetic anisotropy and interfacial Dzyaloshinskii-Moriya interaction in as grown and annealed X/Co/Y ultrathin systems. *J. Phys.: Condens. Matter* **2020**, *32*, No. 495802.
(49) Ourdani, D.; Roussigné, Y.; Chérif, S. M.; Gabor, M. S.; Belmeguenai, M. Dependence of the interfacial Dzyaloshinskii-Moriya interaction, perpendicular magnetic anisotropy, and damping in Co-based systems on the thickness of Pt and Ir layers. *Phys. Rev. B* **2021**, *104*, No. 104421.
(50) Tokaç, M.; Bunyaev, S. A.; Kakazei, G. N.; Schmool, D. S.; Atkinson, D.; Hindmarch, A. T. Interfacial Structure Dependent Spin Mixing Conductance in Cobalt Thin Films. *Phys. Rev. Lett.* **2015**, *115*, No. 056601.
(51) Yanagihara, H.; Kita, E.; Salamon, M. B. Aperiodical oscillation of interlayer coupling in epitaxial co/ir(001) superlattices. *Physical Review B - Condensed Matter and Materials Physics* **1999**, *60*, 12957.
(52) Yang, H. X.; Chshiev, M.; Dieny, B.; Lee, J. H.; Manchon, A.; Shin, K. H. First-principles investigation of the very large perpendicular magnetic anisotropy at Fe—MgO and Co—MgO interfaces. *Physical Review B - Condensed Matter and Materials Physics* **2011**, *84*, No. 054401.
(53) Chung, W. S.; Lee, J. B.; An, G. G.; Yang, S. M.; Kim, J. H.; Hong, J. P. Thermally enhanced perpendicular magnetic anisotropy behaviors of ultrathin [Co/Pd]n multilayers via NiOx capping layer. *Appl. Phys. Lett.* **2015**, *106*, No. 222404.
(54) Chung, W. S.; Yang, S. M.; Kim, T. W.; Hong, J. P. Ultrathin Co-O oxide layer-driven perpendicular magnetic anisotropy in a CoO/[Co/Pd] m multilayer matrix upon annealing. *Sci. Rep.* **2016**, *6*, 37503.
(55) Petitto, S. C.; Marsh, E. M.; Carson, G. A.; Langell, M. A. Cobalt oxide surface chemistry: The interaction of CoO(1 0 0), Co3O4(1 1 0) and Co3O4(1 1 1) with oxygen and water. *J. Mol. Catal. A: Chem.* **2008**, *281*, 49−58.
(56) Pan, C.; An, H.; Harumoto, T.; Zhang, Z.; Nakamura, Y.; Shi, J. Control of the perpendicular magnetic anisotropy and perpendicular exchange bias in CoPt/CoOx thin films. *J. Magn. Magn. Mater.* **2019**, *484*, 320−323.
(57) Moon, K. W.; Lee, J. C.; Choe, S. B.; Shin, K. H. Determination of perpendicular magnetic anisotropy in ultrathin ferromagnetic films by extraordinary Hall voltage measurement. *Rev. Sci. Instrum.* **2009**, *80*, No. 113904.
(58) Barla, A.; Nicolas, J.; Cocco, D.; Valvidares, S. M.; Herrero-Martin, J.; Gargiani, P.; Moldes, J.; Ruget, C.; Pellegrin, E.; Ferrer, S. Design and performance of BOREAS, the beamline for resonant X-ray absorption and scattering experiments at the ALBA synchrotron light source. *J. Synchrotron Rad.* **2016**, *23*, 1507−1517.